\documentclass[12pt]{article}
\usepackage{graphicx}
\usepackage{amsmath,amssymb,bm}
\usepackage{subfig}
\usepackage[square,comma,numbers,sort&compress]{natbib}
\numberwithin{equation}{section}

\newcommand{\ra}{\right\rangle}
\newcommand{\la}{\left\langle}
\newcommand{\hn}{\hat{\mathbf{n}}}
\newcommand{\lb}{\left(}
\newcommand{\rb}{\right)}
\newcommand{\bl}{\mathbf{l}}
\newcommand{\stn}{\frac{\text{S}}{\text{N}}}
\newcommand{\intl}{\int \frac{\mathrm{d}\bl}{(2\pi)^2}}
\newcommand{\intlp}{\int \frac{\mathrm{d}\bl '}{(2\pi)^2}}

\newcommand{\als}{single lens }

\newcommand{\AAA}{{\cal A}}

\newcommand{\be}{\begin{equation}}
\newcommand{\ee}{\end{equation}}
\newcommand{\ben}{\begin{eqnarray}}
\newcommand{\een}{\end{eqnarray}}

\newcommand{\lcdm}{\Lambda\text{CDM}}

\newcommand{\nl}{\hspace{-.65cm}}
\newcommand{\sectiono}[1]{\section{#1}}
\newcommand{\bd}{\begin{displaymath}}
\newcommand{\ed}{\end{displaymath}}
\long\def\symbolfootnote[#1]#2{\begingroup\def\thefootnote{\fnsymbol{footnote}}
\footnote[#1]{#2}\endgroup}

\begin{document}

 \begin{center}

  \textbf{\Large How Sensitive is the CMB to a Single Lens?}

\vspace{10mm}

\normalsize{Ben Rathaus\!\!\symbolfootnote[1]{ben.rathaus@gmail.com}, Anastasia Fialkov\!\!\symbolfootnote[2]{anastasia.fialkov@gmail.com} and Nissan Itzhaki\!\!\symbolfootnote[3] {nitzhaki@post.tau.ac.il}\\\vspace{2mm}{\em Raymond and Beverly Sackler Faculty of Exact Sciences,\\\vspace{1mm}
School of Physics and Astronomy, Tel-Aviv University,\\\vspace{1mm} Ramat-Aviv, 69978,
Israel}}
\end{center}
\vspace{7mm}
\begin{abstract}

\medskip

We study the imprints of a single lens, that breaks statistical isotropy,  on the CMB and calculate the signal to noise ratio (S/N) for its detection. We emphasize the role of non-Gaussianities induced by $\lcdm$ weak lensing in this calculation and show that typically the S/N is much smaller than expected. In particular we find that the hypothesis that a void (texture) is responsible for the WMAP cold spot can barely (cannot) be tested via weak lensing of the CMB.

\end{abstract}

\baselineskip=18pt

\newpage

\section{Introduction }

The role weak lensing plays in CMB physics is well understood in models that assume statistical isotropy,  such as $\lcdm$ \cite{Hu:2000, Lewis:2006, Hanson:2009, Amblard:2004ih}.
Statistical isotropy assures that no preferred direction exists, and weak lensing acts to smear the peaks and troughs of the power spectrum by convolving different scales.

There are, however, some unexpected features in the current data that, at least naively, appear to hint that statistical isotropy is broken at the 2-3$\sigma$ confidence level. These include the drop in the power spectrum at large scales that is highly sensitive to the galactic plane mask \cite{Copi:2006tu, Copi:2010na, Hajian:2007pi}, the alignment of the low $l$ modes \cite{Tegmark:2003ve, Schwarz:2004gk, Land:2005ad, Land:2006bn},  the anomalously large bulk flow \cite{Watkins:2008hf, Feldman:2009es, Lavaux:2008th},  the cold spot \cite{Cruz:2004ce, Cruz:2006sv, Cruz:2006fy, Cruz:2007pe, Cruz:2008sb} and  giant rings \cite{Kovetz:2010kv} (for recent reviews see \cite{Bennett:2010, Abramo:2010, Copi:2010na}).
Cross-correlating these features with  weak lensing  of the CMB could provide us with some valuable information about the origin of  these features.
In particular, it was argued that a possible cosmological explanation to  some of these features could be  an  anomalously large structure (e.g. \cite{Kovetz:2010kv,  Cruz:2007pe,   Cruz:2008sb, Itzhaki:2008ih, Turok:1990gw, Inoue:2006, Fialkov:2009xm}). Such a structure, irrespective of its origin  (which could be new physics or  a statistical fluke within $\lcdm$),   may also leave a statistically significant imprint on the CMB via weak lensing.

With this motivation in mind we study the CMB weak lensing signal associated with a single lens that breaks statistical isotropy. Our main goal is to determine under what circumstances will such a lens leave a statistically significant imprint on the CMB, that is, a  S/N larger than $1$ (for previous works on the subject see \cite{Das:2008, Masina:2009wt,Masina:2010dc}).

The paper is organized as follows. In section 2 we summarize some known results about S/N calculations in two cases that are relevant for weak lensing of the CMB, assuming Gaussian distribution. In section 3 we first show that there is an upper limit on the  S/N associated with a single lens. We refer to this upper limit as  the ideal S/N. We also  calculate the realistic S/N from the temperature CMB data. Section  4 begins with a puzzle: we show that the expressions we got for the realistic S/N imply that an experiment with a resolution $l_{\text{max}}>2000$ has a larger S/N than the ideal S/N. We show that non-Gaussianities induced by $\lcdm$ weak lensing resolve this puzzle in an interesting way. Finally, in section 5 we calculate the S/N  for
two particular examples of large scale structure: a texture and a void. Both are possible cosmological explanations to the cold spot. We find that a texture cannot be detected via weak lensing, whereas a void is barely detectable.

\section{Preliminaries}
\label{sec:prel}

In this section we review two deformations of a Gaussian distribution  and calculate the associated S/N.
As we shall see in the next sections both deformations are relevant for weak lensing of the CMB.

The starting point is a system of $n$ Gaussian fields with vanishing mean values
\[ \langle x^i \rangle =0, ~~~~i=1..n,\]
and non-trivial covariance matrix
\[ \langle x^i x^j \rangle =\text{C}_0^{ij}. \]
The likelihood function associated with this system is \cite{Heavens:2009nx, Hamilton:2005kz, Verde:2009tu}
\be\label{eq:likelihood}
{\cal L}_0= \frac{1}{(2\pi)^{n/2}\sqrt{\det{\text{C}_0}}} \exp{ \left(-\frac12 \mathbf{x}^{\text{T}} \text{C}_0^{-1} \mathbf{x}\right)}.
\ee
For simplicity we take $\text{C}_0$ to be a diagonal matrix.

Generally speaking, there are two classes of deformations of this system that do not induce non-Gaussianity. Since these deformations are from one Gaussian system to another Gaussian system   the S/N associated with them can be calculated exactly.

\vspace{0.3in}
\nl {\it Class 1: Deformation of the mean value}

\nl The Gaussian fields  are shifted by  {\em constants} (i.e. non-random variables)
\be
x^i \rightarrow \tilde{x}^i=x^i+ b^i.
\ee
This deformation modifies the mean values while leaving the covariance matrix intact
\be
\langle \tilde{x}^i \rangle =b^i,~~~~~~~~\langle( \tilde{x}^i - \langle \tilde{x}^i \rangle) (  \tilde{x}_j -  \langle \tilde{x}^j \rangle ) \rangle =\text{C}_0^{ij}.
\ee
The  likelihood function of the deformed system is
\be
{\cal L}(\mathbf{b})= \frac{1}{(2\pi)^{n/2}\sqrt{\det{\text{C}_0}}} \exp{ \left(-\frac12 (\mathbf{x}+\mathbf{b})^{\text{T}} \text{C}_0^{-1} (\mathbf{x}+\mathbf{b})\right)}.
\ee
To calculate the S/N associated with this deformation we proceed in the familiar way, using Fisher information theory (e.g. \cite{Heavens:2009nx, Hamilton:2005kz, Verde:2009tu}). We define
\be
\delta (\mathbf{b})\equiv
-2\left[\log({\cal L}(\mathbf{b}))-\log({\cal L}_0)\right]
=  \left( \mathbf{x}^{\text{T}} \text{C}_0^{-1} \mathbf{b}+ \mathbf{b}^{\text{T}} \text{C}_0^{-1} \mathbf{x} +\mathbf{b}^{\text{T}} \text{C}_0^{-1} \mathbf{b}\right),
\ee
which is nothing but the $\chi^2$ \cite{Heavens:2009nx, Verde:2009tu}, and calculate its mean value which is the desired S/N ratio
\be\label{eq:ideal}
\left(\stn\right)^2 (\mathbf{b})\equiv\langle \delta  (\mathbf{b})\rangle=\int \mathrm{d}x \delta(\mathbf{b})  {\cal L}_0=
\mathbf{b}^{\text{T}} \text{C}_0^{-1} \mathbf{b}.
\ee
This class of deformations is relevant when searching for a known template, $\mathbf{b}$,  in the data.

\vspace{0.3in}
\nl {\it Class 2: Deformation of the covariance matrix}

We leave the mean values intact, $ \langle x^i \rangle =0$, and deform the covariance matrix
\be
\text{C}_0 \rightarrow \text{C}.
\ee
In general, the deformed covariance matrix $\text{C}$ is non-diagonal.
The relevant likelihood function
reads
\be
{\cal L}(\text{C})= \frac{1}{(2\pi)^{n/2}\sqrt{\det{\text{C}}}} \exp{ \left(-\frac12 \mathbf{x}^{\text{T}} \text{C}^{-1} \mathbf{x}\right)}.
\ee
To calculate the S/N we proceed in the same fashion as before.
We define
\be
\delta (\text{C})\equiv -2\left[\log({\cal L}(\text{C}))-\log({\cal L}_0)\right]
=\mathbf{x}^{\text{T}} \left( \text{C}^{-1} -\text{C}_0^{-1}\right) \mathbf{x} +\log\left(\det \text{C} / \det \text{C}_0 \right),
\ee
and calculate its mean value
\be\label{eq:SN2}
\left(\stn\right)^2 \equiv \langle \delta  (\text{C})\rangle=\int \mathrm{d}x \delta(\text{C})  {\cal L}_0= \text{Tr} \left( \text{C}_0 \text{C}^{-1} -  1 \right) +\log\left(\det \text{C} / \det \text{C}_0 \right).
\ee

As is often the case,
weak lensing included, one can expand $\text{C}$ in some small expansion parameter $\epsilon$ as
\be\label{eq:cExpand}
\text{C}=\text{C}_0+\epsilon \text{C}_1+\frac{\epsilon^2}{2} C_2 +...
\ee
Expanding (\ref{eq:SN2}) in $\epsilon$ we find that the linear term vanishes
due to cancellation between the two terms in the r.h.s. of  (\ref{eq:SN2}). This follows from the fact that  $\text{C}=\text{C}_0$ is a local minimum. What is somewhat surprising is that the leading term, that scales as $\epsilon^2$, depends only on $\text{C}_1$ (and not  on $\text{C}_2$)
\be\label{1}
\left(\stn\right)^2=\frac{\epsilon^2}{2} \sum_{i j} \frac{ |\text{C}_1^{ij}|^2 }{\text{C}_0^{ii} \text{C}_0^{jj}}.
\ee

A familiar situation, which is irrelevant to our case, is that of diagonal $\text{C}_1$. Then the leading S/N reads
\be
\left( \stn \right)^2 = \frac{1}{2}\sum_{i} \left( \frac{\delta \text{C}_{ii}}{\text{C}^0_{ii}} \right )^2,
\ee
with $\delta \text{C}_{ii} = \epsilon C_1$.

Applying this result to the Gaussian statistically isotropic  CMB temperature (where $i$ runs over $l$ and $m$) we get the familiar expression
\be
\left( \stn \right)^2 = \sum_{l} \left( \frac{\delta \text{C}_{l}}{\Delta \text{C}^0_{l}} \right )^2,\ee
where $ \Delta \text{C}^0_l=\text{C}_l\sqrt{2/(2l+1)} $.

\sectiono{Ideal and Realistic Expressions for S/N}

In this section we apply the deformations discussed above to calculate the S/N associated with weak lensing of the CMB induced by a single lens.
We perform two calculations. One is of an ideal S/N in which the signal of the weak lensing  by the \als  is competing only with the noise associated with $\lcdm$ weak lensing. The second is a realistic S/N in which the signal is extracted from the temperature data. In such a case, the signal is competing with various contributions to the noise on top of the noise associated with $\lcdm$ weak lensing. Thus we expect the ideal S/N to serve as an upper bound on the realistic S/N.

\subsection{Ideal S/N}

In $\lcdm$ the deflection potential field $\psi^\Lambda (\hn)$
is statistically isotropic and follows a Gaussian distribution \cite{Hu:2000, Lewis:2006, Hanson:2009}
\be \label{eq:psiStatistics}
\langle \psi_{l m}^{\Lambda} \rangle =0,~~~~~\langle {\psi^{\Lambda}}_{lm}^* \psi^{\Lambda}_{l'm'} \rangle =\delta_{l l'}\delta_{m m'} C^{\psi}_l,
\ee
where as usual we expand $\psi^\Lambda(\hn) = \sum_{lm}\psi_{lm}Y_{lm}(\hn)$ and the angle brackets stand for an ensemble average.

The effect of a \als on $\psi_{l m}^{\Lambda}$ is to shift it by a constant $ \delta\psi_{lm}$ that depends on the details of the \als
\be\label{eq:psiShift}
\psi^{\Lambda}_{l m} \to \psi^{\Lambda}_{l m} +\delta\psi_{l m}.
\ee
Since the shift is by a constant (and not a random field) the deformation associated with adding a \als to $\lcdm$ is of the first class of deformations discussed in the previous section, and the ideal S/N is
\be \label{eq:idelFull}
\left(\stn\right)^2_{\text{Ideal}}=\sum_{l m} \frac{|\delta\psi_{l m}|^2}{ C^{\psi}_l}.
\ee

In the rest of the paper we work mostly in the flat-sky approximation. The full- and flat-sky variables are related in the following way \cite{Hu:2000}
\be
C(\bl) = C_l,~~~~~~\psi(\bl) = \sqrt{\frac{4\pi}{2l+1}} \sum_m i^{-m} \psi_{lm} e^{im\phi_l}.
\ee
In flat-sky variables (\ref{eq:psiStatistics}) takes the form
\be
\la \psi^{\Lambda}(\bl) \ra =0,~~~~~\la {\psi^{\Lambda}}^*(\bl) \psi^{\Lambda}(\bl ') \ra =(2\pi)^2 \delta(\bl -\bl ')C^{\psi}(\bl),
\ee
and the ideal S/N is
\be\label{eq:idealS2n}
\left(\stn\right)^2_{\text{Ideal}}=\intl \frac{\left|\delta\psi(\bl)\right|^2}{ C^{\psi}(\bl)},
\ee
where $\bl$ is a 2D momentum.

\subsection{Realistic S/N}

The effect of the \als on the CMB temperature is to shift the temperature  at each point in a way that is determined by the deflection potential it induces, $\delta\psi$. Expanding to leading orders we have \cite{Lewis:2006, Hanson:2009, Amblard:2004ih}
\be\label{rq}
  T^{\text{SL}}(\hn) = T(\hn)+\nabla^i\delta\psi\nabla_i T(\hn)+\frac{1}{2}\nabla^i\delta\psi\nabla^j\delta\psi\nabla_i\nabla_j T(\hn)\dots
\ee
where $T$ is the $\lcdm-$lensed temperature field and $T^{\text{SL}}$  is further lensed by the single lens.

To calculate the S/N it is natural  to define
$
\delta T(\hn) =  T^{\text{SL}}(\hn) - T(\hn)
$
and treat it as a deformation of the first class.
In that case we can use (\ref{eq:ideal}), to find
\be
\left(\stn\right)^2_{\text{Temp}}=\la \delta T^{\text{T}} \text{C}^{-1} \delta T \ra.
\ee
This is the point of view taken in \cite{Das:2008} which in the flat-sky approximation yields for a spherically symmetric single lens\footnote{With the substitution $\intl \rightarrow \sum_{l} \frac{2l+1}{2}$.}
\be\label{ds}
\left(\stn\right)^2_{\text{Temp}}=\sum_{l} \frac{2l+1}{2} \frac{\tilde{S}_l}{C_l},
\ee
where
\be\label{eq:sLDas}
\tilde{S}_l=\intlp \left|\bl '\cdot (\bl-\bl ')\delta\psi(\bl ')\right|^2 C(\left|\bl-\bl ' \right|).
\ee
This seems correct since a known structure determines
$\delta\psi$ and so it determines $\delta T(\hn)$. Hence it appears that  once again we are searching for a known template, this time $\delta T(\hn)$, in the data.

However, as emphasized in the previous section this approach can be used only when the template is a known {\em constant} deformation  and not a random field. In our case $\delta\psi$ is a constant deformation but $\delta T(\hn)$ is a random field. In particular, its mean value vanishes $
\langle \delta T(\hn) \rangle =0.$ We conclude that from the point of view of temperature fluctuations  weak lensing of the CMB by a \als is not a deformation of the first class. Hence (\ref{eq:ideal}) is not a good starting point for the calculation of the S/N.
In fact,  it is a deformation of the second class.

To see this and calculate the S/N it is convenient to
Fourier transform  (\ref{rq}) and get the familiar result \cite{Lewis:2006, Hanson:2009}
\ben \label{eq:tSL}
T^{\text{SL}}(\bl) &=& T(\bl) - \intlp \bl ' \cdot (\bl - \bl ')\delta\psi(\bl ') T(\bl - \bl ') \nonumber\\
&& - \frac{1}{2} \intlp \frac{\mathrm{d}\bl ''}{(2\pi)^2} \bl ' \cdot \left[\bl ' + \bl '' - \bl \right] \bl ' \cdot \bl '' T(\bl ') \delta\psi(\bl '') \delta\psi^{*}(\bl '+\bl '' - \bl).
\een
It is easy to see that indeed $\la  T^{\text{SL}}(\bl) \ra=0$ and that
the covariance matrix
\be
\text{Cov}(\bl_1,\bl_2) \equiv \la {T^{\text{SL}}}^*(\bl_1)   T^{\text{SL}}(\bl_2)  \ra,
\ee
is deformed.  Eq. (\ref{1}) implies that to calculate the S/N to second order in  $\delta\psi$ we need to know  $\text{Cov}(\bl_1,\bl_2)$  only to first order. This is easy to calculate. For  convenience we first define
\be\label{eq:gammaDef}
\gamma(\bl_1,\bl_2) \equiv \left( \bl_1+\bl_2\right)\cdot \left[ \bl_1 C(\bl_1)+\bl_2 C(\bl_2) \right] 
\ee
so that the off-diagonal terms of the covariance matrix are
\be
\text{Cov}(\bl_1,\bl_2) \underset{\bl_1\neq -\bl_2}{=} \gamma(-\bl_1,\bl_2)\delta\psi(\bl_2-\bl_1).
\ee
The diagonal terms of the deformation vanish\footnote{This statement is correct only in the flat-sky approximation. In full-sky, there is a small diagonal contribution  to the covariance matrix, as we show in appendix \ref{sec:fullSky}.}
so that the diagonal terms of the complete covariance matrix are just the original $C(l)$'s.

Thus using (\ref{1}) we find
\be
\label{eq:FlatSN2}\left(\stn\right)^2_{\text{Temp}}
=\frac{1}{2}\intl \frac{\mathrm{d}\bl '}{(2\pi)^2}  \frac{\left| \gamma(-\bl,\bl ')\delta\psi(\bl '-\bl )\right|^2 }{C(l)C(l ')}.
\ee
In the case of a spherically symmetric deflection potential,
for which $\delta\psi(\bl) = \delta\psi(l),$ one can write (\ref{eq:FlatSN2}) as
\be\label{eq:s2nTemSl}
\left(\stn\right)^2_{\text{Temp}}= \sum_{l} \frac{2l+1}{2}\frac{S_l}{C_l},
\ee
where
\be \label{eq:sL}
S_l \equiv \intlp \frac{\left|\delta\psi(l ')\right|^2 }{2C(l ')}\left[ \bl '\cdot \lb \bl\, C(l) + (\bl '-\bl ) C(\left|\bl '-\bl  \right|) \rb
\right]^2.
\ee
Note that while  Eq. (\ref{eq:sL}) looks similar to  Eq. (\ref{eq:sLDas}), because of the cancelation between the two terms in (\ref{eq:sL}) it is in fact smaller
by roughly a factor of $10^{3}$. This makes a single lens much harder to detect via weak lensing of the CMB than expected.

\section{The Role of Non-Gaussianities}
\label{sec:nonG}

In the previous section we have calculated both the ideal and realistic S/N associated with a single lens. As we now show it turns out that the realistic S/N is larger than the ideal S/N for experiments, such as Planck, SPT and ACT, with 6' resolution or better. This puzzle is illustrated in the context of a toy example in the next subsection. In the second subsection we show how
non-Gaussianities induced by weak lensing of the CMB resolve this puzzle.

\subsection{The Toy-Model and the Puzzle}

So far we have had two types of signal to noise ratios: the ideal (\ref{eq:idealS2n}), and the realistic (\ref{eq:FlatSN2}).
We have argued that the ideal S/N must be  an upper bound on any realistic S/N, and in particular on (\ref{eq:FlatSN2}). This should hold for any single lens.
Consider for simplicity a  toy-model in which all the contributions to the deflection potential comes from a single mode $\bl_0$
\be\label{eq:toyDef}
\delta\psi(\bl)=(2\pi)^2 \AAA\, \delta(\bl - \bl_0).
\ee
The ideal S/N is then simply
\be
\left(\stn\right)^2_{\text{Ideal}}= \frac{\AAA \, \delta\psi(\bl_0)} {C^{\psi}(l_0)},
\ee
and the realistic S/N is
\be\label{eq:toyS2N}
\left(\stn\right)^2_{\text{Temp}}= \frac{\AAA\,\delta\psi(\bl_0)}{2}\intl
\frac{\left| \gamma(\bl,\bl_0-\bl) \right|^2}
{C(l)C(\left| \bl_0-\bl \right|)}.
\ee
In light of the above argument we expect the ratio
\be\label{eq:RDef}
R(l_0) \equiv \frac{\left(\text{S}/\text{N}\right)^2_{\text{Temp}}} {\left(\text{S}/\text{N}\right)^2_{\text{Ideal}}}= \frac{C^{\psi}(l_0)}{2}\intl
\frac{\left| \gamma(\bl,\bl_0-\bl) \right|^2}
{C(l)C(\left| \bl_0-\bl \right|)}
\ee
to be smaller than 1 for all $l_0$.

\begin{figure}
\begin{center}
\subfloat[]{\label{fig:cross}\includegraphics[width=.46\textwidth] {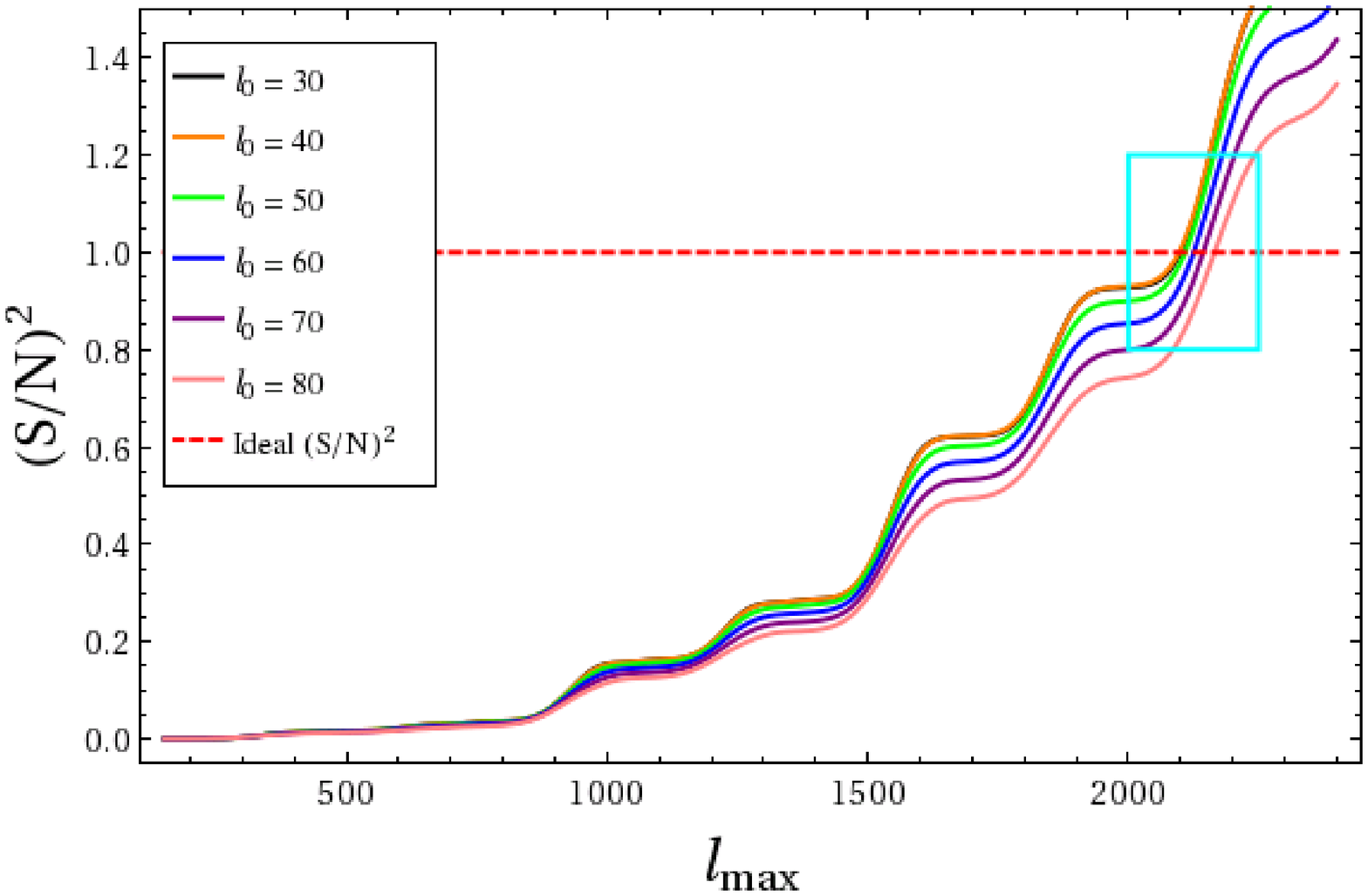}}
\qquad \subfloat[]{\label{fig:zoom}\includegraphics[width=.48\textwidth]{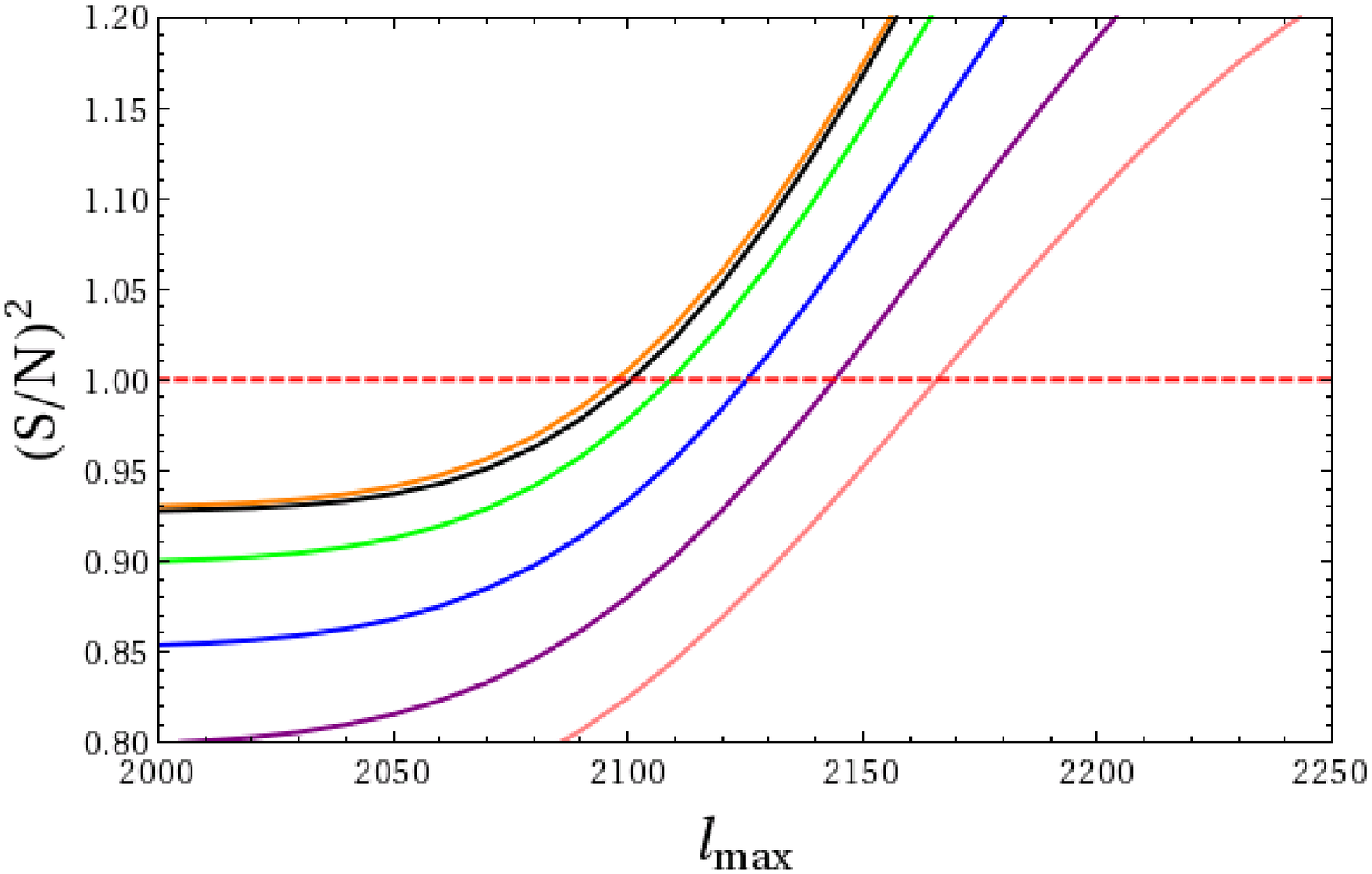}}
\caption{$\text{S}/\text{N}_{\text{Temp}}$ for the toy model $\delta\psi \propto \delta(\bl-\bl_0)$, as described in (\ref{eq:toyDef}) through (\ref{eq:RDef}) for a few values of $l_0$, normalized according to the $\text{S}/\text{N}_{\text{Ideal}}$ (red dashed line). Here \protect\subref{fig:cross} shows the cumulative S/N as a function of $L_{\text{max}}$ for a wide range of $L_{\text{max}}$ while \protect\subref{fig:zoom} is a zoom-in
picture
of the cyan rectangle, emphasizing the region of $L_{\max}$ where S/N$_{\text{Temp}}$ crosses S/N$_{\text{Ideal}}$.}
\label{fig:crossover}
\end{center}
\end{figure}

Writing the integral in (\ref{eq:toyS2N}) as
\be
\intl~~~ \rightarrow~~~ \int \frac{\mathrm{d}\phi}{(2\pi)^2} \int_{0}^{L_{\text{max}}} l\mathrm{d}l
\ee
quantifies the resolution capabilities of a given experiment via the cutoff $L_{\text{max}}$.\footnote{Alternatively one can add the instrumental noise $N_l\sim \exp(l^2/l_{max}^2)$.} Calculating $\text{S}/\text{N}_{\text{Temp}}$ we find that for \emph{any} $l_0$ it scales quadratically with $L_{\text{max}}$. Thus $R(l_0)$ crosses $1$  for a
sufficiently large $L_{\text{max}}$, which is typically of order 2000 (see Fig. \ref{fig:crossover}). The quadratic dependence on $L_{\text{max}}$ follows from the fact that at large $l$ the dependence of $(\text{S}/\text{N})_{\text{Temp}}$ on $C(l)$ drops and we are left with an integral over a constant. Therefore, this puzzle is generic and does not depend on the details of the model and/or parameters of $\lcdm$.

\subsection{Non-Gaussianities Resolve this Issue}

As is often the case with puzzles, the origin of this puzzle is hidden in the assumptions we have made. In deriving (\ref{1}) we assumed  that the background temperature field follows a Gaussian distribution. However, it is well known that weak lensing induces non-Gaussianity \cite{Lewis:2006, Hu:2004}.  In $\lcdm$  this non-Gaussianity is expected to start affecting the Gaussian behavior at $l\sim 1000$. At $l\sim 2000$ it modifies the Gaussian results by about $5\%$ \cite{Hu:2004}.
In fact, very recently this effect was measured \cite{Smidt:2010by, Das:2011ak}, and an agrement with theory was found.
As we show below,  when statistical isotropy is broken, the non-Gaussianity induced by weak lensing is more significant: It alters our findings by a factor of order
unity exactly where it should, $l\sim 2000$, in order to solve the puzzle we have encountered.

\begin{figure}
\begin{center}
\subfloat[]{\label{fig:oneL}\includegraphics[width=.25\textwidth]{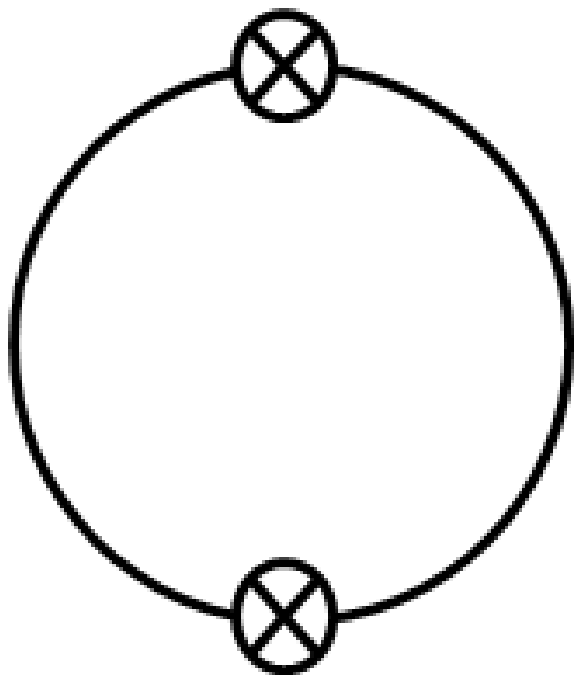}}
\quad\quad\subfloat[]{\label{fig:twoL}\includegraphics[width=.4\textwidth]{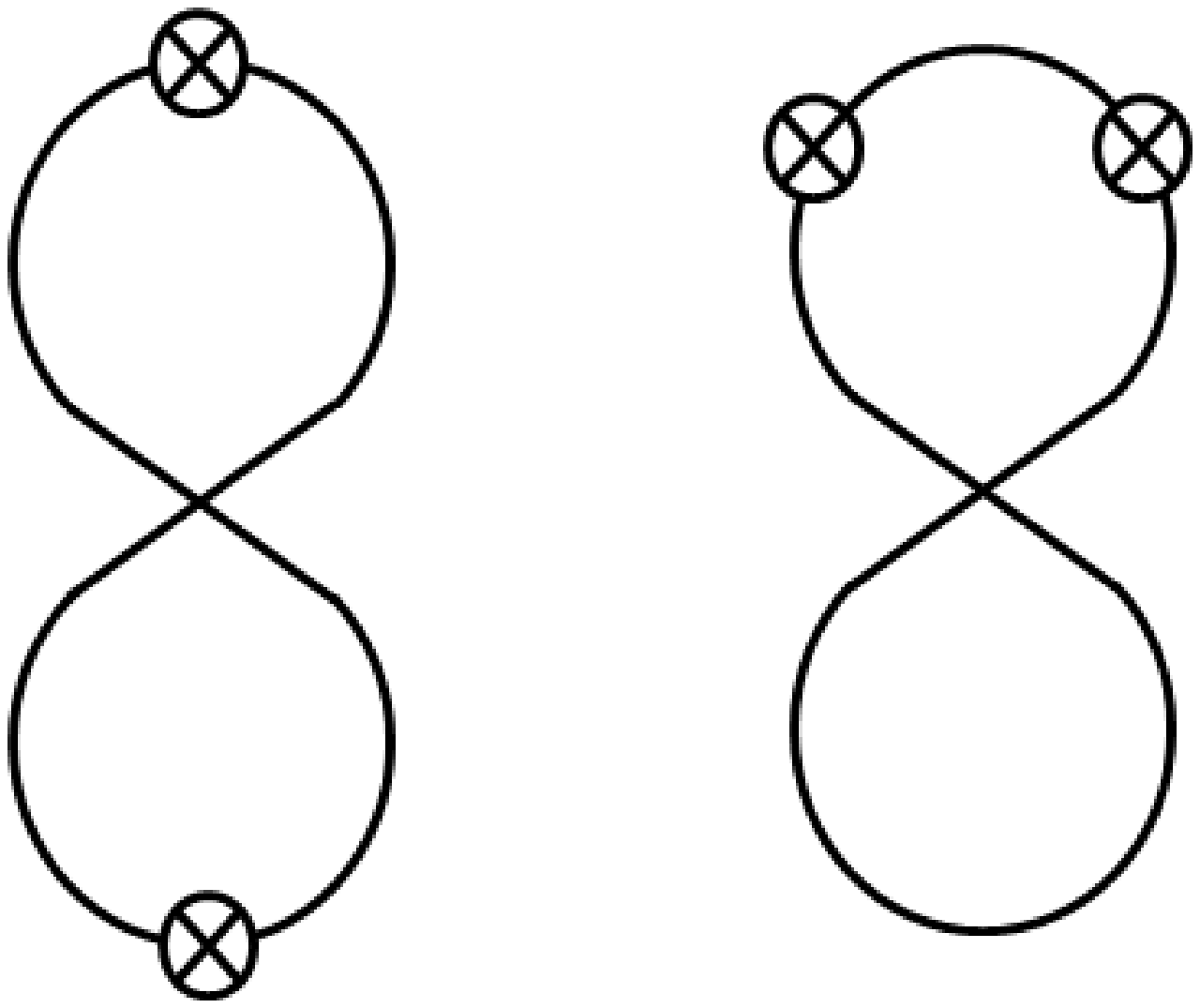}}
\caption{Visualizing the S/N$_{\text{Temp}}$ as vacuum-energy Feynmann diagrams up to 2-loop order following the Feynman rules of section \ref{sec:nonG}. The diagrams here are schematic, and do not account for some complication due to $\lcdm$ four-leg vertex substructure. \protect\subref{fig:oneL} The one-loop (Gaussian) part of the S/N. \protect\subref{fig:twoL} The leading non-Gaussian contribution.}
\label{fig:split2Loop}
\end{center}
\end{figure}

It is constructive to address this calculation 
via Feynmann diagrams.
Let us first repeat the S/N calculation in the Gaussian case of the previous section using Feynmann diagrams.
The Feymann rules associated with this Gaussian theory are as follows:
\begin{itemize}
\item Assign a 2D momentum to each leg in the diagram, and a propagator, $C(l)$, which corresponds to the $\Lambda$CDM (lensed) temperature power spectrum.
\item  The single lens induces a two-leg ``interaction'' vertex $\tilde{\gamma}(\bl_1,\bl_2)$/2, where  $\tilde{\gamma}(\bl_1,\bl_2)= \gamma(\bl_1,\bl_2)\delta\psi(\bl_1+\bl_2)/C(l_1)C(l_2)$
\begin{center}{\includegraphics[width=.35\textwidth]{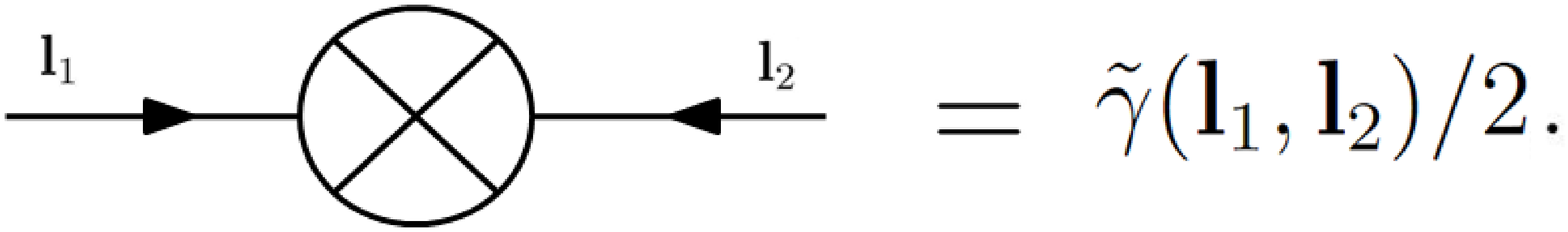}} \end{center}
This vertex mixes between two momentum modes.
\item Multiply each diagram by a proper symmetry factor. The relevant symmetry factors, and their derivation are elaborated on in Appendix \ref{sec:feynmannRules}.
\end{itemize}
The Gaussian calculation of the previous section is presented by  the diagram of Fig. \ref{fig:oneL}. Namely, it is a vacuum energy diagram with an external background field (that breaks statistical isotropy) induced by the single lens. Following the Feynmann rules we get, using (\ref{eq:gammaDef}) and the fact that $C(l)$ and $\psi(\hn)$ are real,
\be\label{eq:s2nWithFeyn}
\lb\stn\rb_{\text{Temp}}^{2} =  \frac{1}{2}\int \frac{\mathrm{d}\bl_1}{(2\pi)^2} \frac{\mathrm{d}\bl_2}{(2\pi)^2}C(l_1)C(l_2) \left|\tilde{\gamma}(-\bl_1,\bl_2)\right|^2,
\ee
which is identical to (\ref{eq:FlatSN2}).

$\lcdm$ weak lensing induces non-Gaussianities in the temperature field. Assuming  negligible  correlations between the deflection potential and the primordial temperature the leading non-Gaussianity is a four-leg vertex. Hence we have to supplement our Feynmann rules with
\begin{itemize}
\item $\lcdm$ weak lensing induces a four-leg vertex
\begin{center}{\includegraphics[width=.35\textwidth]{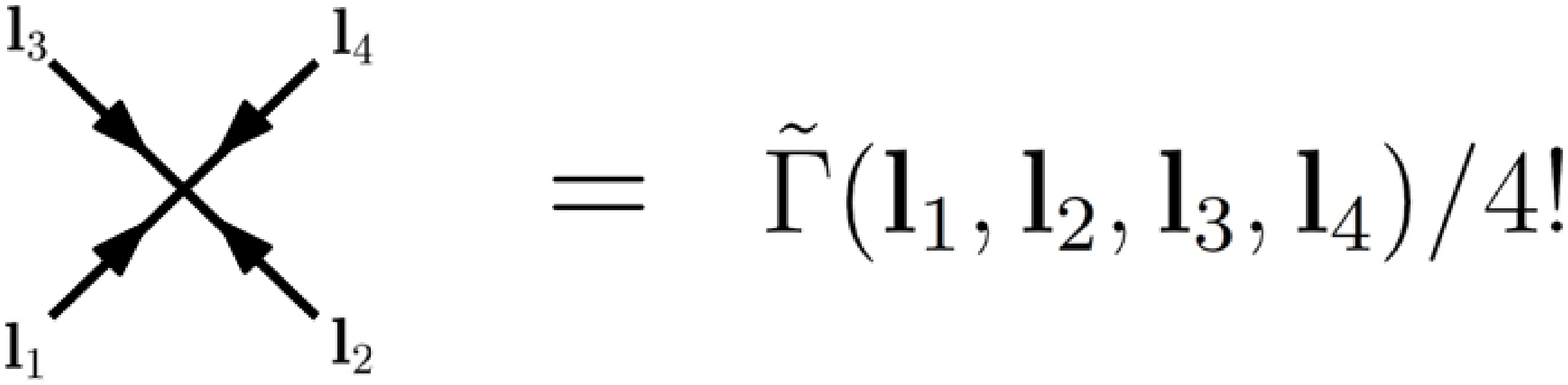}} \end{center}

\end{itemize}

\begin{figure}
\begin{center}
\subfloat[]{\label{fig:perL}\includegraphics[width=.48\textwidth]{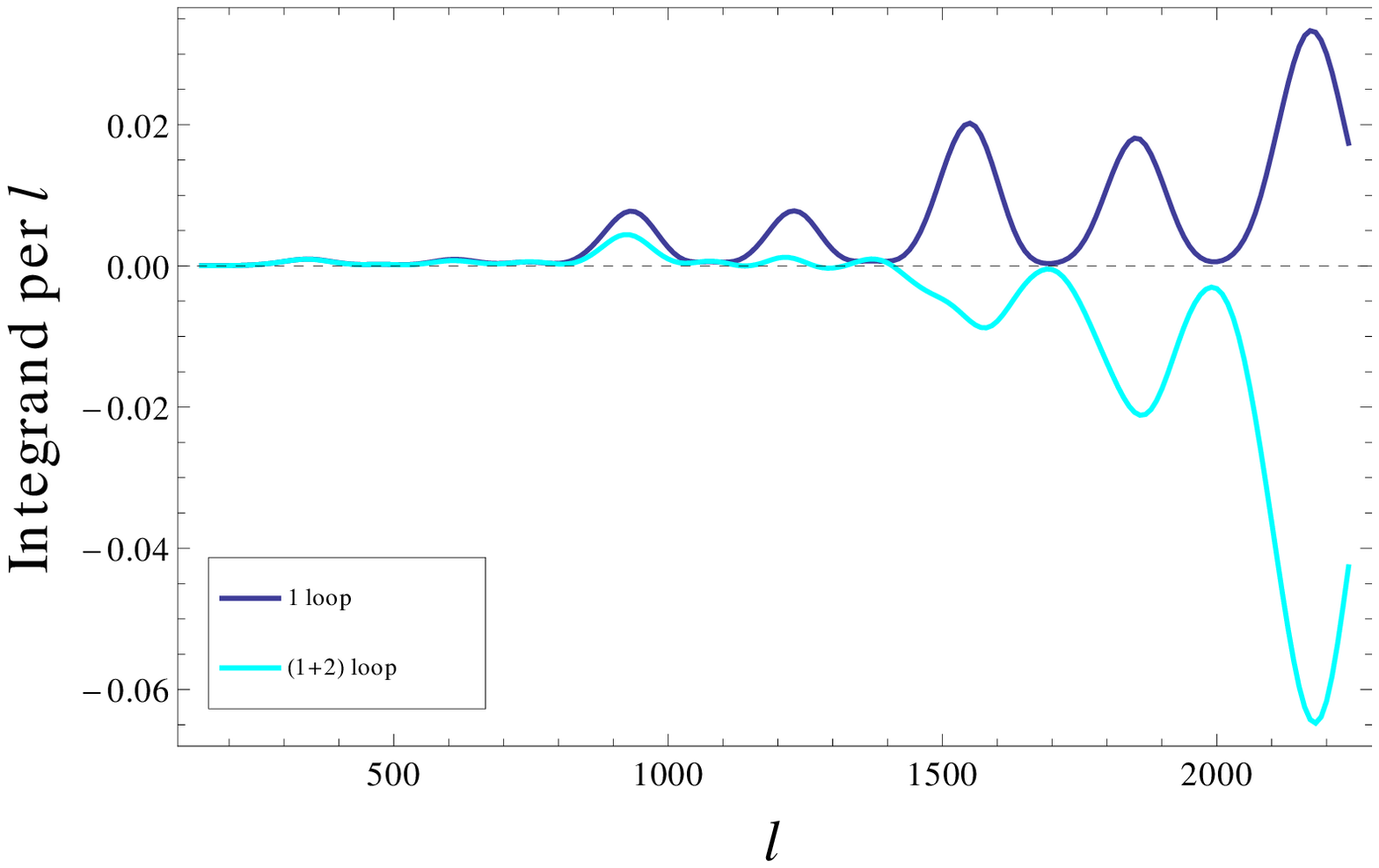}}
\subfloat[]{\label{fig:totalAcc} \includegraphics[width=.48\textwidth]{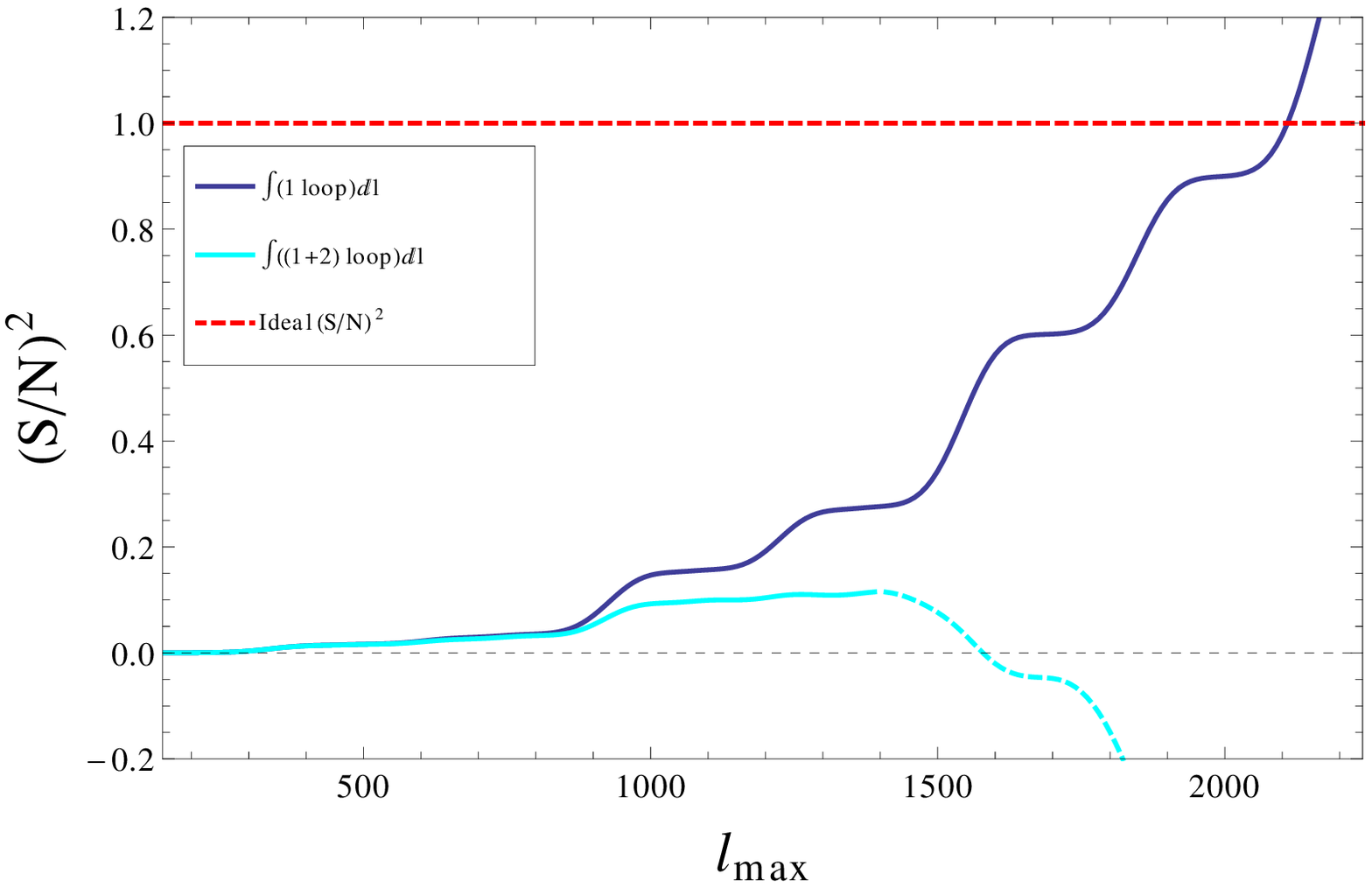}}
\caption{The significance of the 2-loop contribution to the S/N$_{\text{Temp}}$. Here \protect\subref{fig:perL} shows the 1-loop (blue) vs. $(1+2)-$loop (cyan) contribution to the \emph{integrand} per $l$, while  \protect\subref{fig:totalAcc} shows the $(\text{S}/\text{N})_{\text{Ideal}}^{2}$ (red dashed), Gaussian part of $(\text{S}/\text{N})_{\text{Temp}}^{2}$ (blue solid), and $(\text{S}/\text{N})_{\text{Ideal}}^{2}$ (cyan solid and dashed) to leading non-Gaussian contribution for $l_0=50$. The cyan line becomes dashed at the point where the overall non-Gassian contribution to the integrand becomes negative.}
\label{fig:accumulated1and2loop}
\end{center}
\end{figure}

Since the contributions to the S/N we are interested in are quadratic in $\delta\psi$, the relevant leading\footnote{Leading here refers to the number of loops in a diagram or, alternatively,  the number of times the 4-leg vertex appears in a diagram.} Feynmann diagrams are those of Fig \ref{fig:twoL}.
In standard field theories (e.g. $\phi^4$) such diagrams are easy to calculate. Here, however, the vertex is rather complicated. The reason is that it has  substructure that follows from the relation between the unlensed and the lensed correlation functions in $\lcdm$, leading to topologically distinct diagrams. The details of the calculation are left to Appendix \ref{sec:feynmannRules}. Here we merely quote the results. There are four different diagrams which contribute to the 2-loop S/N$_{\text{Temp}}$. One of these is positive, and the other three are negative. The sign of the overall leading non-Gaussian contribution to $\text{S}/\text{N}_{\text{Temp}}$ is \emph{negative}. Therefore, it lowers $(\text{S}/\text{N})_{\text{Temp}}^{2}$, as is illustrated in Fig. \ref{fig:accumulated1and2loop}.

Fig. \ref{fig:accumulated1and2loop} also shows that the leading non-Gaussian contribution becomes significant at $l\sim 900$ and that for $1000<l_{\text{max}}<1500$ there is an approximated plateau in the accumulated S/N. The value of $(\text{S/N})^2$ at this plateau is about 1/10 of the ideal $(\text{S/N})^2$. These features are not sensitive to $l_0$. For $l_{\text{max}}>1500$ the accumulated S/N starts to drop. This is a nonphysical artifact of  the fact that we keep only the first non-Gaussian correction. We expect higher (in loop counting sense) non-Gaussian corrections to remove this artifact and to alter the S/N for $l_{\text{max}}>1500$. It is beyond the scope of this paper to calculate these corrections and to determine the exact dependence of $\text{S}/\text{N}_{\text{Temp}}$ on $l_{\text{max}}$. We emphasize  that, irrespective of $l_{\text{max}}$,   one cannot exceed $\sim \frac{1}{\sqrt{10}} \text{S}/\text{N}_{\text{Ideal}}$ using  the temperature data alone without  performing this higher loop calculation.

Physical situations, like the ones discussed in the next section,  can be viewed  as a superposition of toy models with different $l_0$'s. Therefore we expect them to behave in a similar way, namely
\be\label{eq:s2nApprox}
\lb \stn\rb _{\text{Temp}}\sim \frac{1}{\sqrt{10}} \lb\stn\rb_{\text{Ideal}}
\ee
to be a good approximation.

\section{Physical Examples}

In this section we study the weak lensing S/N associated with two examples of spherically symmetric structures that were proposed
as possible cosmological explanations to the WMAP cold spot \cite{ Cruz:2007pe, Cruz:2008sb, Das:2008, Masina:2009wt}.
One is  a cosmic texture \cite{Turok:1990gw, Cruz:2007pe, Das:2008} and the other is a localized  void (similar to that discussed in \cite{Inoue:2006, Das:2008, Masina:2009wt}).

\subsection{Cosmic Texture}
The deflection profile of a texture as discussed in \cite{Das:2008}, is
\be
\alpha(\theta)=A_T \frac{\theta}{\sqrt{1+4(\theta/R_T)^2}},
\ee
which is valid in the range $0\leq \theta <R_T$. We follow \cite{Das:2008}, and take $R_T = 5^{\circ}$ and $A_T = 5.19 \times 10^{-4}$ to best describe the WMAP cold spot. In the flat-sky approximation we have \cite{Das:2008}
\be
\delta\psi(l) = -\frac{\pi}{2}A_TR_T e^{-lR_T/2}(lR_T+2)/l^3,
\ee
from which we can calculate the ideal S/N to find (using (\ref{eq:idealS2n}))
\be
\lb \stn\rb_{\text{Ideal}}^{2}=f_{\text{sky}}~ 2 \times 10^{-2},
\ee
where $f_{sky}$ is the observed fraction of the sky. We conclude that the texture's imprint on the CMB via weak lensing is too small to be detected.

\subsection{Void}

The void case is somewhat more elaborate because there is a subtlety in the ideal S/N calculation that should be dealt with.
Our starting point is the void discussed in \cite{Das:2008} which was taken to have a {\em sharp} wall at  $R_V = 3.1^{\circ}$. Namely, the energy density of the void is
\be
\frac{\delta\rho(\theta)}{\rho_0}=\left \{ \begin{array}{l l}
0~ &\theta\leq R_V\\ 1~ &\theta>R_V ~.
\end{array}
\right.
\ee
The deflection angle associated with such a void is
\be \label{eq:voidalpha}
\alpha_V(\theta)=\left \{ \begin{array}{l l}A_V \theta&\theta\leq R_V\\ A_V\frac{R_V^2}{\theta}&\theta>R_V
\end{array}
\right.
\ee
with $ A_V=0.01785$.
We see that the discontinuity in $\delta \rho$ induces  a  discontinuity in the first derivative of $\alpha_V(\theta)$ at $\theta=R_V$.

\begin{figure}
\begin{center}
\includegraphics[width=.6\textwidth]{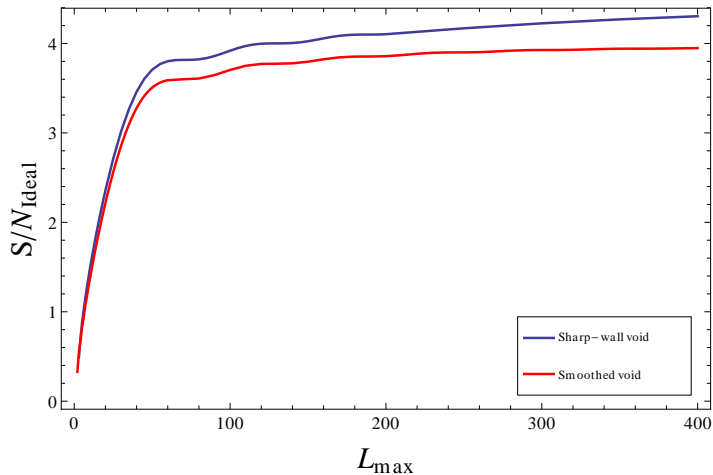}
\caption{The ideal S/N of the void, in two ``versions'': a void which is modelled by a deflection profile with a sharp wall (blue) as in  (\ref{eq:voidalpha}) and which is seen to moderately diverge with $L_{\text{max}}$, and a void whose profile was smoothed so that the first derivative of the deflection profile is continuous (red), which converges.}
\label{fig:s2nvoid}
\end{center}
\end{figure}

Upon Fourier transforming (\ref{eq:voidalpha}), this discontinuity in position space translates to a signal in momentum space which does not decay fast enough for the ideal S/N to converge.  Indeed, at large $l$, the deflection potential as in \cite{Das:2008}
\be
\delta \psi= 4\pi A_V R_V J_1(l R_V)/l^3,
\ee
scales like $\sim l^{-7/2}$ while the deflection potential power spectrum of $\lcdm$ drops faster than $l^{-4}$ so the S/N of the void  diverges as anticipated (see Fig. \ref{fig:s2nvoid}).
This is a rather mild divergence (only logarithmic in $l$) that is not going to affect any calculation  relevant for actual experiments. Still, this issue is worth addressing.

The origin of this divergence is the assumption that the void is spherically symmetric with a sharp wall. In physical situations it is often the case that the thickness  of the wall is much smaller than the size of the void and so a sharp wall is a good approximation. However, we do not expect voids to be approximately spherically symmetric. Hence a discontinuity in the first derivative of  $\alpha_V(\theta)$ is very unlikely to happen.
We have considered various smoothings  of this discontinuity.
Since the original divergence is only logarithmic in $l$ we do not expect the details of the smoothing to considerably affect  the S/N. Indeed, this is what we have found with a typical value of
\be
\lb\stn\rb _{\text{Ideal}}^{2} \sim 15 \, f_{\text{sky}}.
\ee
Hence using (\ref{eq:s2nApprox}) the realistic S/N is approximately
\be
\lb\stn\rb  _{\text{Temp}}\sim  1.25\,\sqrt{f_{\text{sky}}}.
\ee
We conclude that in experiments with partial sky coverage and high angular resolution, such as ACT and SPT, $f_{\text{sky}}$ renders the void undetectable. Moreover, even in full-sky high resolution experiments such a void is barely detectable.

\vspace{4mm}

\noindent {\bf Acknowledgements}

\vspace{4mm}

We thank A. Ben-David, E. Kovetz and M. Shimon for discussions. This work is
supported in part by the Israel Science Foundation (grant number
1362/08) and by the European Research Council (grant number 203247).

\vspace{4mm}

\appendix
\sectiono{Full-sky S/N} \label{sec:fullSky}

The full-sky expression for the S/N in the case of a spherically symmetric lens reads \be \label{eq:s2nSymb}
\lb\frac{\textrm{S}}{\textrm{N}}\rb _{\text{Temp}}^{2} = \sum_{l\leq l_2,m} \frac{ \lb C_{lm;l_2m}^{(1)}\rb^{2}}{C_lC_{l_2}}\lb 1-\frac{\delta_{ll_2}}{2} \rb,
\ee
where $C_{lm;l_2m_2}^{(1)}$ is the leading order correction to the covariance matrix, that is,
\be \label{eq:clml2m2}
C_{lm;l_2m_2}^{(1)} =  \la T_{lm} ^{*} a_{l_2m_2}^{(1)}  \ra +  \la {a_{lm}^{(1)}}^{*} T_{l_2m_2} \ra,
\ee
and where $T_{lm}$ and $a_{lm}^{(1)}$ are the $l,m-$harmonic moments of the $\lcdm$ lensed temperature field, and its first order correction due to the single lens respectively.

The expression (\ref{eq:s2nSymb}) includes both diagonal and off-diagonal contributions,
and implies that the covariance matrix is diagonal in the ``$m-$dimension'' (since we consider a spherically symmetric lens).
The off-diagonal terms $C_{lm;l_2m}^{(1)}$ (with $l\neq l_2$), are derived in \cite{Masina:2010dc} (Eq. 13), and here we mainly examine the diagonal contribution to the S/N (\ref{eq:s2nSymb}).

\subsection{Diagonal Terms of the Covariance Matrix}

The diagonal terms are obtained  by taking $l=l_2$ in (\ref{eq:clml2m2}) (or in Eq. (13) of \cite{Masina:2010dc}). This gives
\ben \label{eq:diagCov}
C_{lm;lm}^{(1)}   &=&  C_l\sum _{l''}  (-1)^{m}  l''(l''+1)  \nonumber\\
&&~~~~~~~~~\times \lb \begin{array}{ccc}
l&l&l''\\0&0&0
\end{array}\rb \lb \begin{array}{ccc}
l&l&l''\\-m&m&0
\end{array}\rb \nonumber\\
&&~~~~~~~~~\times(2l+1)\sqrt{\frac{2l''+1}{4\pi}}
\delta \psi_{l''0} ,
\een
where
\[ \lb \begin{array}{ccc}
l&l'&l''\\m&m'&m''
\end{array}\rb \]
is the Wigner-$3j$ symbol, and where we have fixed the coordinate system so that only $m=0$ modes of the deflection potential are non-vanishing.
Hence,
\be
\sum_{m}  \lb C_{lm;lm}^{(1)}\rb^{2}= \sum_{ l''} \frac{(2l+1)^2}{4\pi} \left[ \lb \begin{array}{ccc}
l&l&l''\\0&0&0
\end{array}\rb \right]^2  \left[l''(l''+1)\right]^2 C_{l}^{2}\delta\psi_{l''0}^2.
\ee
We see that, unlike in the flat-sky approximation, the full-sky expression contains a non-vanishing contribution.

Despite the fact that there is a diagonal contribution to S/N at first order we show now that the first order contribution to the \emph{estimator} of the lensed power-spectrum vanishes. To see this note that since
\be
\hat{\tilde{C}}_{l}=\sum_{m} \frac{C_{lm;lm}}{2l+1},
\ee
where $\hat{X}$ denotes the estimator of $X$, and since
\be
 \sum_m (-1)^m \lb \begin{array}{ccc}
l&l&l''\\-m&m&0
\end{array}\rb \propto \delta_{l''0}.
\ee
we get
\be
\hat{\tilde{C}}_{l}^{(1)} = 0,
\ee
as claimed.

\subsection{S/N}

Collecting both contributions, the diagonal of the previous section and off-diagonal calculated in \cite{Masina:2010dc}, we find that
\be
\lb\frac{\textrm{S}}{\textrm{N}}\rb_{\text{Temp}} ^{2} = \lb\frac{\textrm{S}}{\textrm{N}}\rb _{\textrm{DP}}^{2} + \lb\frac{\textrm{S}}{\textrm{N}}\rb _{\textrm{ODP}}^{2},
\ee
where
\be
\lb\frac{\textrm{S}}{\textrm{N}}\rb _{\textrm{DP}}^{2} = \sum_{lm} \frac{ \lb C_{lm;lm}^{(1)}\rb^{2}}{2C_{l}^{2}},
\ee
and where
\ben
\lb\frac{\textrm{S}}{\textrm{N}}\rb _{\textrm{ODP}}^{2} &=& \frac{1}{4}\sum_{l < l_2, l''} \frac{(2l+1)(2l_2+1)}{4\pi C_lC_{l_2}} \left[ \lb \begin{array}{ccc}
l&l_2&l''\\0&0&0
\end{array}\rb \right]^2\nonumber\\
&&~~~~~~~~~~\times  \lb\left[ l(l+1)-l_2(l_2+1) + l''(l''+1) \right] C_{l}   \right. \nonumber\\
 && ~~~~~~~~~~~~~~ + \left. \left[ l_2(l_2+1)-l(l+1) + l''(l''+1) \right] C_{l_2}   \rb^2 \delta\psi_{l''0}^2.
\een

\section{Performing 2-Loop Calculations}
\label{sec:feynmannRules}
In section \ref{sec:nonG} we have already outlined the Feynmann rules, which make the 2-loop calculation easier. In this appendix we
write down explicitly the ``vertex-functions'' $\tilde{\gamma}$ and $\tilde{\Gamma}$, draw all the diagrams that contribute to the S/N up to 2-loop order, find their symmetry factors, and then use the Feynmann rules to evaluate the different contributions. We then apply the toy-model deflection (\ref{eq:toyDef}) to the results, thus simplifying them further. These results give together the 2-loop negative contribution seen in Fig. \ref{fig:accumulated1and2loop}. All the calculations in this appendix are carried out in the flat-sky approximation.

\subsection{The Vertex-Functions}
A vertex-function is the correlation function (in Fourier-space) divided by the incoming propagators. Therefore the 2-leg vertex function of a \als is
\ben \label{eq:tildegamma}
\tilde{\gamma}(\bl_1,\bl_2)&\equiv& \frac{\la T^{\text{SL}}(\bl_1) T^{\text{SL}}(\bl_2)\ra}{C(l_1)C(l_2)}\nonumber\\
&=& \frac{\gamma(\bl_1,\bl_2) \delta\psi(\bl_1+\bl_2)} {C(l_1)C(l_2)} \nonumber\\
&\underset{\bl_1\neq -\bl_2}{=}   & \lb\bl_1 +\bl_2\rb \cdot \left[ \bl_1  C(l_1)+\bl_2 C(l_2) \right]   \frac{\delta\psi\lb \bl_1+\bl_2\rb}{C(l_1)C(l_2)} +\mathcal{O}\lb \delta\psi^2\rb.
\een
The 4-leg vertex induced by the $\lcdm$-weak-lensing effects is more complicated since it has non-trivial substructure. This follows from the fact that the vertex interaction of the lensed temperature field is calculated from the interaction of the unlensed temperature field and $\psi^{\Lambda}$,
namely,
\begin{center}{\includegraphics[width=.6\textwidth]{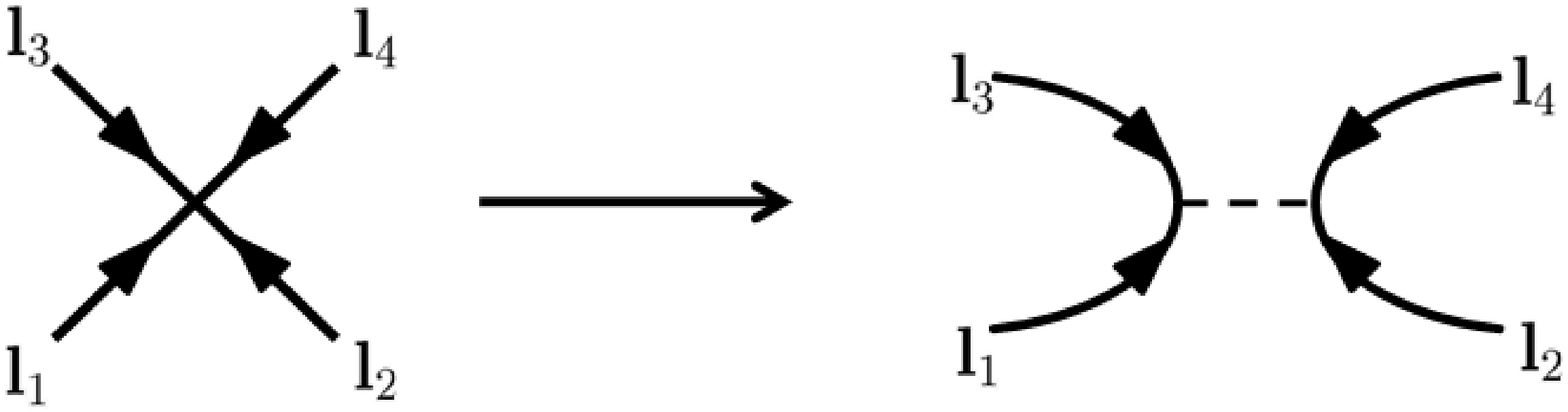} }\end{center}
where, all the legs on the l.h.s. are of the lensed temperature, and on the r.h.s. they are of the unlensed temperatures. The dashed line is the $\psi^\Lambda$ propagator, $C^\psi$.

For the above drawn orientation of the momenta the correlation function may be written as
\begin{multline} \label{eq:fourPnt}
\la  T (\mathbf{l}_1)  T (\mathbf{l}_2)  T (\mathbf{l}_3)  T (\mathbf{l}_4) \ra_{\text{drawn orientation}} \equiv \\
\lb 2\pi \rb ^2 \delta(\mathbf{l}_1 +\mathbf{l}_2 +\mathbf{l}_3 +\mathbf{l}_4)  \, \, \mathbf{l}_3 \cdot\lb \mathbf{l}_1+\mathbf{l}_3 \rb \mathbf{l}_4 \cdot\lb \mathbf{l}_2+\mathbf{l}_4 \rb C^{\scriptstyle{\text{unlen}}}(l_3)C^{\scriptstyle{\text{unlen}}} (l_4)C^{\psi}(\left|\mathbf{l}_1+\mathbf{l}_3\right|).
\end{multline}
This statement implicitly \emph{assigns a different role to each leg in the diagram}, so each permutation of $\{ \bl _1, \bl_2, \bl_3, \bl_4 \}$ (there are 24), represents in principle a different vertex.

By analogy to (\ref{eq:tildegamma}) the ``4-point vertex-function'' of section \ref{sec:nonG} is
\be
\tilde{\Gamma}(\bl _1, \bl_2, \bl_3, \bl_4) = \frac{\la  T (\mathbf{l}_1)  T (\mathbf{l}_2)  T (\mathbf{l}_3)  T (\mathbf{l}_4) \ra}{C(l_1)C(l_2)C(l_3)C(l_4)}.
\ee
Here $\la  T (\mathbf{l}_1)  T (\mathbf{l}_2)  T (\mathbf{l}_3)  T (\mathbf{l}_4) \ra$ includes all the permutations and $C(l)$ are  incoming ($\Lambda$CDM lensed) propagators.
\subsection{Closing the Diagrams and Symmetry Factors}

The Feynmann rules for the $\tilde\gamma$ and $\tilde\Gamma$ vertices include, as usual,  division by $2$ and $4!$ respectively. This follows from the number of possible permutations of the legs which enter the vertex, and is standard in real field theories.
The symmetry factor of a diagram is the number of different possible ways to generate it. For example since the diagram of Fig. \ref{fig:oneL} may be generated in two different ways one gets the $(1/2)^2 \times 2=1/2$ in front of the integral (\ref{eq:s2nWithFeyn}).

For the 2-loop calculation we have to draw all possible diagrams which correspond to Fig. \ref{fig:twoL}. For this purpose we have to form loops of the four legs of the $\tilde{\Gamma}$ vertex, and then add the two $\tilde{\gamma}$ vertices in all possible ways. For each diagram of the 24 permutations, there are three ways one can form a vaccum energy diagram. These are
\begin{center}{\includegraphics[width=.6\textwidth]{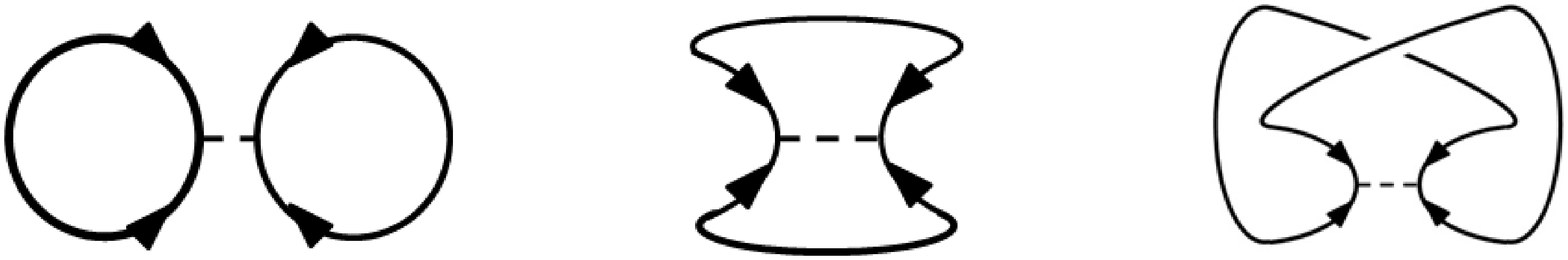} }\end{center}

The left diagram is 1-particle reducible, so we ignore it.
For each of the other two diagrams there are three ways one can insert the two $\tilde{\gamma}$ vertices (each of which can be inserted in two ways, which cancels the $1/2$ that comes with $\tilde{\gamma}$). We further note that two of the three ways produce topologically identical diagrams, such as
\begin{center} {\includegraphics[width=.4\textwidth]{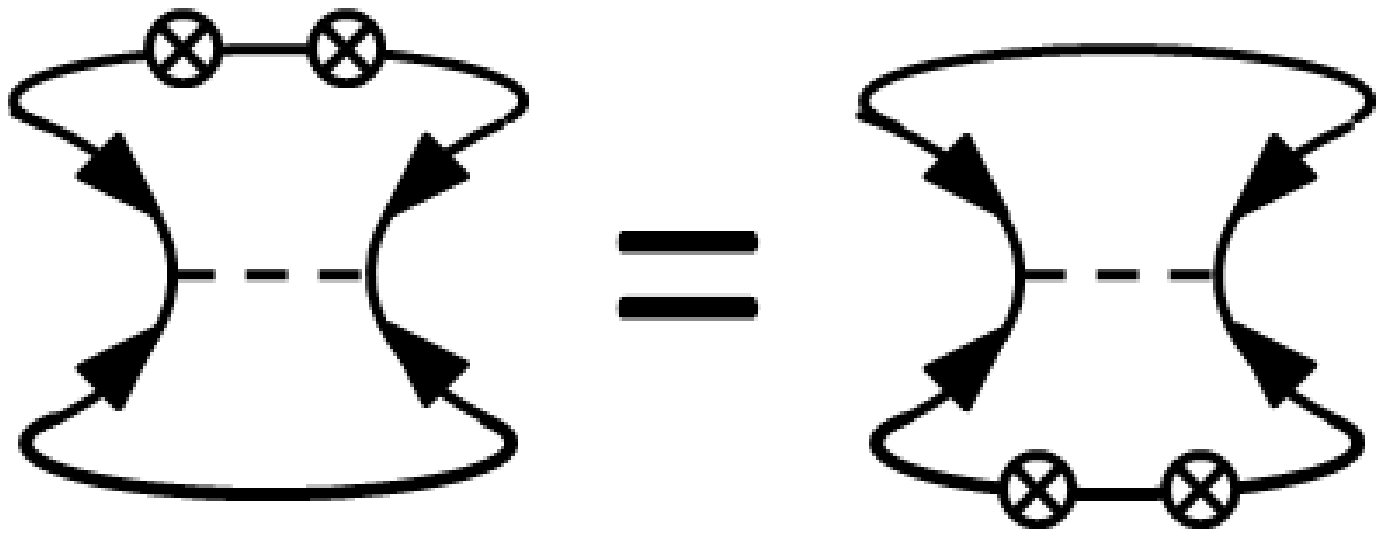} }\end{center}
and therefore all the distinct contributions may be described as
\begin{center} {\includegraphics[width=.5\textwidth]{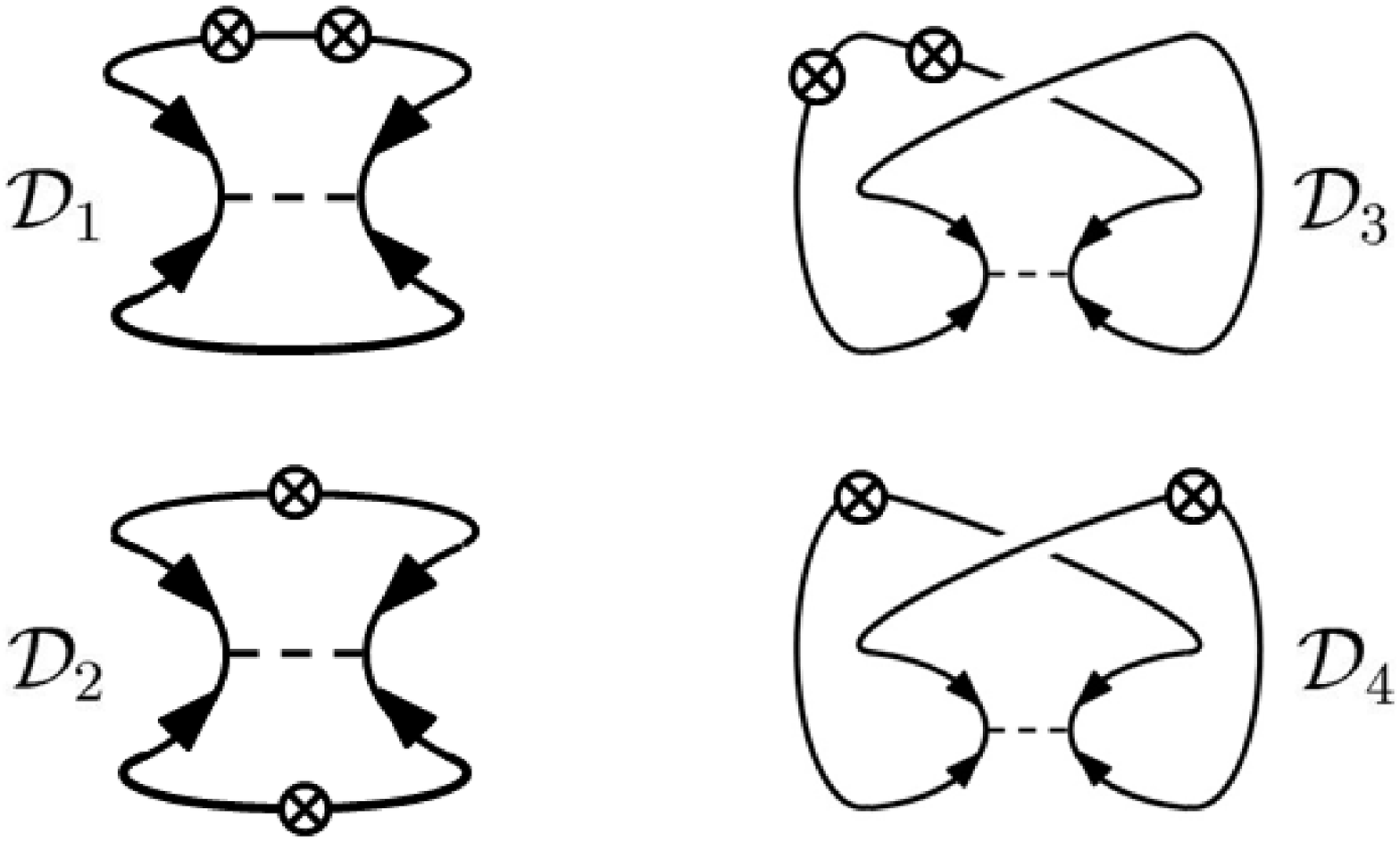} }\end{center}
where we have named the different diagrams, and where each diagram in the top row actually represents 2 separate diagrams.

Therefore, the weights of the diagrams (symmetry factor divided by a factor according to the Feynmann rules) are
\be \label{eq:weight}
\begin{array}{ccccl}
48/4!& = & 2& & \text{for each diagram in the top row}\\
24/4!& = & 1& & \text{for each diagram in the bottom row.}
 \end{array}
\ee

\subsection{Evaluating the 2-Loop Diagrams}

We now turn to evaluating the 1-particle irreducible 2-loop diagrams.
We have
\ben
\mathcal D _1&\equiv& w_1 \int\prod_{i=1}^{4}\frac{\mathrm{d}\bl_i} {(2\pi)^2}  C(l_1)  C(l_3) C(l_4)  C(l_2) \tilde{\Gamma} (\bl _1, -\bl_1, \bl_3, \bl_4) \tilde \gamma(-\bl_3,-\bl_2) \tilde\gamma(\bl_2,-\bl_4) \\
\mathcal D _2 &\equiv& w_2\int\prod_{i=1}^{4}\frac{\mathrm{d}\bl_i}{(2\pi)^2}  C(l_1) C(l_2) C(l_3) C(l_4) \tilde\Gamma (\bl _1, \bl_2, \bl_3, \bl_4) \tilde \gamma(-\bl_1,-\bl_2)
\tilde\gamma(-\bl_3,-\bl_4)\\
\mathcal D _3 &\equiv&  w_3\int\prod_{i=1}^{4}\frac{\mathrm{d}\bl_i}{(2\pi)^2}  C(l_1)  C(l_2) C(l_3)  C(l_4) \tilde\Gamma (\bl _1, \bl_2, \bl_3, -\bl_1) \tilde\gamma(-\bl_3,-\bl_4)
\tilde\gamma(\bl_4,-\bl_2) \\
\mathcal D _4 &\equiv& w_4 \int\prod_{i=1}^{4}\frac{\mathrm{d}\bl_i}{(2\pi)^2}  C(l_1) C(l_2) C(l_3) C(l_4)   \tilde\Gamma (\bl _1, \bl_2, \bl_3, \bl_4) \tilde\gamma(-\bl_2,-\bl_3)
\tilde\gamma(-\bl_1,-\bl_4)
\een
respectively, where $w_i$ is the corresponding weight, according to (\ref{eq:weight}).

\begin{figure}
\begin{center}
\includegraphics[width=.57\textwidth]{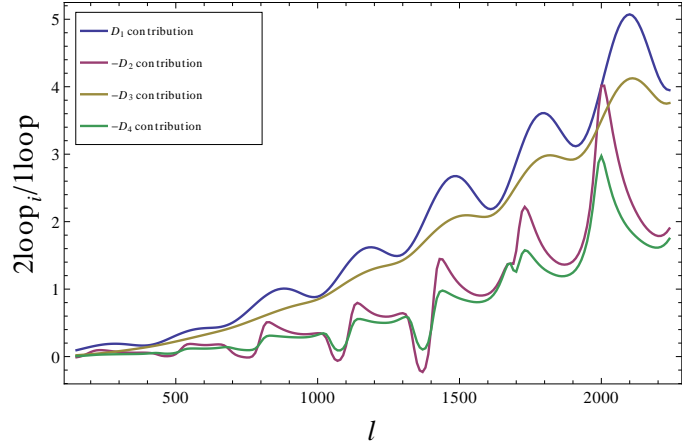}
\caption{The four separate non-Gaussian contributions ($\mathcal{D}_1,\,\mathcal{D}_2, \,\mathcal{D}_3$ and $\mathcal{D}_4$) relative to the 1-loop contribution to the S/N$_{\text{Temp}}$, for the toy model discussed in section \ref{sec:nonG}, per $l$. As may be read from the graph, the $\mathcal{D}_1$ contribution is positive, while the other three contributions are negative. }
\label{fig:twoOverOne}
\end{center}
\end{figure}

Applying the toy model (\ref{eq:toyDef}) gives the four separate contributions which may be seen  in Fig. \ref{fig:twoOverOne}. Simplifying the expressions as much as possible one gets
\ben
\mathcal D _1&=&w_1 \AAA\,\, \delta\psi  \lb \bl_0\rb \int\frac{\mathrm{d}\bl}{(2\pi)^2} \frac{{C^{\scriptstyle{\text{unlen}}}}^2(l)}{C(l)} \frac{\left|\gamma(\bl,\bl_0-\bl)  \right|^2} {C(l)C(\left| \bl_0-\bl\right|  ) } \left[\int\frac{\mathrm{d}\mathbf{k}}{(2\pi)^2}   \frac{\left[ \bl \cdot\mathbf{k}  \right]^2 C^{\psi}(k)}{C(\left|\mathbf{k} - \bl \right|) } \right]\\
\mathcal D _2 &=&- w_2 \AAA \,\delta\psi (\bl_0)\int \frac{\mathrm{d}\bl}{(2\pi)^2} C^{\scriptstyle{\text{unlen}}}(l) C^{\scriptstyle{\text{unlen}}} (\left|\bl_0 + \bl  \right|) \frac{\gamma(\bl,-\bl_0-\bl)}{C(l)C(\left|\bl_0 + \bl  \right|)} \nonumber\\
&&\quad \quad \quad  \quad \quad ~ \left[ \int \frac{\mathrm{d}\mathbf{k}}{(2\pi)^2} \left[  \mathbf{k} \cdot\bl  \right]   \left[\mathbf{k} \cdot  \lb \bl_0+\bl \rb \right] \frac{\gamma(\mathbf{k}-\bl,\bl_0+\bl-\mathbf{k})}{C(\left| \mathbf{k} - \bl\right|)C(\left|  \bl_0-\bl_1 \right|)}C^{\psi}(k) \right] \\
\mathcal D _3 &=& w_3 \AAA\, \delta\psi  \lb \bl_0 \rb \int \frac{\mathrm{d}\bl}{(2\pi)^2} \frac{C^{\scriptstyle{\text{unlen}}} (l) }{C(l)} \frac{\left| \gamma(\bl,\bl_0-\bl)\right|^2} {C(l) C(\left| \bl_0 -\bl\right|)}\nonumber \\
&&\quad \quad \quad \quad \quad \left[ \int \frac{\mathrm{d}\mathbf{k}}{(2\pi)^2}
\left[ \mathbf{k} \cdot \bl \right] \left[ \mathbf{k} \cdot \lb \mathbf{k}-\bl \rb  \right]  \frac{C^{\scriptstyle{\text{unlen}}} (\left| \mathbf{k} - \bl\right|) }{C(\left| \mathbf{k} - \bl\right|)} C^{\psi}(k) \right] \\
\mathcal D _4 &=& w_4 \AAA\,\delta\psi  (\bl_ 0) \int \frac{\mathrm{d}\bl}{(2\pi)^2} C^{\scriptstyle{\text{unlen}}} (l) \frac{\gamma(\bl,-\bl_0-\bl)}{C(l)C(l)} \left[ \int \frac{\mathrm{d}\mathbf{k}}{(2\pi)^2} \left[ \mathbf{k} \cdot \bl \right] \left[ \mathbf{k} \cdot \lb \bl_0 +\bl -\mathbf{k}\rb  \right] \right.\nonumber\\
&& \quad \quad \quad \quad \quad \quad \quad ~~~~~ \left. C^{\scriptstyle{\text{unlen}}} (\left| \bl_0 +\bl -\mathbf{k}\right|) \frac{C^{\psi}(k)\gamma(\mathbf{k}-\bl,\bl_0+\bl-\mathbf{k}) }{C(\left|\mathbf{k} -\bl \right|) C(\left|\bl_0 +\bl -\mathbf{k} \right|)} \right].
\een
The overall contribution of the 2-loop diagrams to the S/N is negative, and yields the plateau in the accumulated S/N$_{\text{Temp}}$ as seen in Fig. \ref{fig:totalAcc}.

\end{document}